%% file: sample-acmtog.tex
\documentclass[acmtog, 12pt, manuscript]{acmart}
\renewcommand\footnotetextcopyrightpermission[1]{} 
\usepackage{booktabs} 

\usepackage[ruled]{algorithm2e} 
\usepackage{breqn}

\SetAlFnt{\small}
\SetAlCapFnt{\small}
\SetAlCapNameFnt{\small}
\SetAlCapHSkip{0pt}
\IncMargin{-\parindent}

\begin{document}
\title{The Skiplist-Based LSM Tree} 
\author{Aron Szanto}
\affiliation{%
  \institution{Harvard University}
  \city{Cambridge}
  \state{MA}
  \postcode{02138}
  \country{USA}}
\setcopyright{none}
\settopmatter{printacmref=false}
\begin{abstract}
Log-Structured Merge (LSM) Trees provide a tiered data storage and retrieval paradigm that is attractive for write-optimized data systems. Maintaining an efficient buffer in memory and deferring updates past their initial write-time, the structure provides quick operations over hot data. Because each layer of the structure is logically separate from the others, the structure is also conducive to opportunistic and granular optimization. In this paper, we introduce the Skiplist-Based LSM Tree (sLSM), a novel system in which the memory buffer of the LSM is composed of a sequence of skiplists. We develop theoretical and experimental results that demonstrate that the breadth of tuning parameters inherent to the sLSM allows it broad flexibility for excellent performance across a wide variety of workloads.
\end{abstract}
\maketitle
\input{samplebody-journals}

\bibliographystyle{ACM-Reference-Format}
\bibliography{bib}

\end{document}

%% file: samplebody-journals.tex
\section{Introduction}
As data scales, transactional updates and reads become more costly. Traditional systems do not differentiate between hot and cold data, foregoing significant optimization opportunities, since in many applications users need to access the most recent data the fastest. The LSM tree, introduced in 1995 by O'Neil et al.\cite{O'Neil96thelog-structured}, provides a mechanism for quick updates, deletes, and writes by collecting them in a pool of active keys before pushing them to secondary storage when the pool is full. By arranging secondary storage in tiers, the cost of merging the buffer to disk is amortized, allowing for efficient writes, while the maintenance of hot keys in memory allows for performant lookups over recent data. The tiered structure of data also provides a natural opportunity for indexing. A variety of indexing structures, including fence pointers, zone maps, and Bloom filters are commonly used to minimize unnecessary disk accesses. In addition, compression algorithms can be used to shrink the memory and disk footprint of the data both in memory and on disk. Because LSM trees have disparate and independent components, there is a large space for optimization. However, the parameters of interaction between the components are also a crucial part of good performance. In this paper, we describe a novel LSM system that uses cache-conscious skiplists in memory, along with Bloom filter and fence pointer indexing, to achieve excellent throughput. The remainder of this paper will proceed as follows: Section 2 will detail the design of the Skiplist-Based LSM (sLSM), including the in-memory component, the on-disk component, indexing structures, key algorithms, theoretical guarantees, and the range of design knobs; Section 3 will provide extensive experimental results, including parameter tuning and performance analysis; and Section 4 will discuss and conclude. 

\section{sLSM Design}
The sLSM has two macroscopic components: the in-memory buffer and the disk-based store. The in-memory section is composed of a set of data structures that is optimized for quick insert and lookup on the buffered data. The disk-based store is composed of a tiered layer storage that scales by a constant factor with each tier.

\subsection{Memory Buffer}
The memory buffer consists of $R$ runs, indexed by $r$. In the sLSM , one run corresponds to one skiplist. Only one run is active at any one time, denoted by the index $r_{a}$. Each run can contain up to $R_n$ elements. An insert occurs as follows: if the current run is full, make the current run a new, empty one. If there is no space for a new run: i) merge a proportion $m$ of the existing runs in the buffer to secondary storage; ii) set the active run to a new, empty one. Insert the key-value pair into the active run. A lookup is similar: starting from the newest run and moving towards the oldest, search the skiplist for the given key. Return the first one found (as it is the newest). If not found, then search disk storage.

\subsection{Skiplists}

Skiplists are probabilistic data structures that provide for fast search within an ordered sequence of values. They are composed of decreasingly sparse sorted runs of values that are set in parallel, so that a search consists of searching a run until a key is found that is greater than the desired one, then repeating the same process on the next-densest run, until the correct key is found. With some careful optimization, these structures can be powerful, yet leave only a small memory footprint. Two optimizations implemented in sLSM are presented below.
\subsubsection{Fast Random Levels}
One of the vital steps in skiplist insertion is choosing the "level" that the element to be inserted will occupy. Ideally, the distribution of levels follows a geometric distribution with parameter $p$; in practice, $p = 0.5$ is standard, and also has nice mathematical properties regarding optimal average runtime  \cite{pugh1990skip}. The sLSM uses hardware optimization to generate random levels quickly. Rather than an iterative mechanism that increments the level with probability $p$ at each round, stopping when the level is not incremented, we propose an $O(1)$ solution: generate $MAXLEVEL$ random bits, where $MAXLEVEL$ is the maximum level for any element. Return the result of the hardware builtin "find first set bit" function (standard on x86-64). Since each bit is random, the probability that the $n$th bit is the first one that is set is $2^{-n}$, which is exactly the geometric distribution with parameter $p = .5$ that is needed. Our skiplists use $MAXLEVEL = 16$, which was experimentally determined to be optimal. There is a tradeoff between the probabilistic speed of retrieval and the skiplist size, including the amount of data needed to be loaded into the cache for a lookup. As $MAXLEVEL$ gets larger, there is a higher probability that nodes can be skipped, leading to faster lookup for higher-valued keys. However, the list of forward pointers is larger, and may not fit in the cache, leading to cache misses that offset the performance gains of high-level nodes. We found that fitting the forward pointer list into two cache lines is optimal, and theorize that this is due to the fact that the second cache line is only accessed about $1-(.5^8) = 0.4\%$ of the time. When it is accessed, the speedup due to skipping large swaths of the list outweighs the infrequent performance drawdown of loading the extra cache line.
\subsubsection{Vertical Arrays, Horizontal Pointers}
\begin{figure}
\centering
\includegraphics[scale=.5]{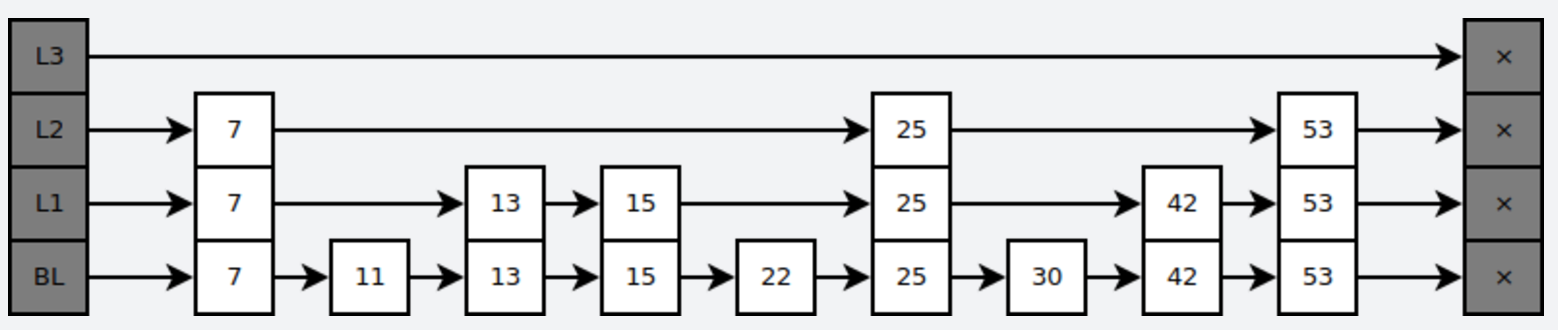}
\caption{}
\label{fig:cacheconsciousSL}
\end{figure}

The other skiplist optimization involves the way that the skiplist traverses levels. While the differential densities of the levels precludes an array-based structure in the horizontal direction, it is wasteful to include links from nodes of value $k$ to another node of value $k$ on the next level. Instead, in our implementation a skiplist node includes one key, one value, and an array of pointers to other skiplist nodes. This array can be thought of as a vertical column of level pointers, where pointers above the node's level are null and each pointer below points to the next node on that particular level. In this way, skipping down a level is a matter of reading a value that was already loaded into the cache, rather than chasing a pointer somewhere random in memory. Because we implemented the skiplist in this way from the beginning we do not report differential performance between this cache-conscious and the alternative, naive way. Fig. \ref{fig:cacheconsciousSL} summarizes this optimization nicely.\footnote{Source: http://ticki.github.io/blog/skip-lists-done-right/ (MIT LICENSE)}\\

\subsection{Memory Buffer Indexing}
Bloom filters are space-efficient probabilistic data structures that are used to test whether an element is in a set. Using a series of hash functions and a bitset, the filter can provide a strong probabilistic guarantee: an element will never induce a false-negative result under a test for membership, and an element will only induce a false-positive result up to some error probability $\epsilon$, a value that is chosen by the user and that is traded off against the space occupied by the filter. Bloom filters find important use in the sLSM when paired one-to-one with runs in memory and on disk. Rather than incur a high cost by searching for a key in every run, the filter is consulted first; if it returns negative, we can safely skip that run, because of the filter's no-false-negative guarantee. In this way, we'll only search a proportion $\epsilon$ of the runs that we don't need to, which could result in a significant time saving for lookups. In our implementation, Bloom filters are leveraged by pairing each consideration of a run with a filter test; if it fails, we simply skip that run. We use the Murmur3 hash function and utilize the mathematical technique of "double hashing", allowing us to quickly generate the $k$ hash values necessary without recomputing the entire hash $k$ times by using a linear combination of two hashes for each. We also keep track of the maximal and minimal key in each run for low-cost, high-granularity filtering by run.
\subsection{Disk-Backed Storage}
The disk-backed store is for more permanent storage and is only touched by merges from the memory buffer. There are $L$ disk levels, each with $D$ runs on each level. A run is an immutable sequence of sorted key-value pairs, and is captured within one memory-mapped file on disk. Levels grow by runs until they hit the threshold $D$, and then a new level is added, incrementing $L$. Each disk run is indexed similarly to the in-memory run, with max/min keys and Bloom filters. Additionally, we use fence pointers to index into disk runs for lookups. These are fixed-width indices that store the key of elements in increments of some logical page size in memory. To look up a key in a disk run, we find the fence pointers that bound the key via binary search, then search for the key in that range on disk, also by binary search. This reduces disk accesses by a factor of $\frac{\log n}{\log \mu}$, where $\mu$ is the fence pointer page size.
\subsection{Merging}
One of the most important implementation details of the sLSM is the merging algorithm. When the buffer becomes full, a fraction $m$ of the runs is flagged and their elements collected and sorted, then written to the shallowest disk level's next available run. In this way, adjacent levels share the following relationship: the size of a level's runs is identical to the total size of its shallower neighbor multiplied by the fraction of runs merged $m$. Analogously, the number of elements at level $k$ is $O((mD)^k)$. Merging is not as simple as copying from file to file, however. A disk level might be full, requiring a cascade of merges down to lower disk levels. This complex operation is quite nuanced: though the runs being merged are individually sorted, the resulting run needs to be sorted. Because runs at lower levels do not fit in memory, some optimizations are necessary to save both time and space. Moreover, when several runs from level $k$ contain the same key, the value that remains tied to that key on level $k + 1$ must be the most recently written, i.e., the one that came from the newest run on level $k$. The naive algorithm for merging $n$ items from $k$ sorted lists is $O(nk)$, where each element is compared against the minimal unwritten item from each list. We propose a heap-based merging algorithm that runs in $O(n \log (mD))$ time and $O(mD)$ space, where $n$ is the number of elements being merged. Algorithm 1 demonstrates our approach, which uses a min-heap whose constituents are pairs of key-value elements and integers denoting from which run that key-value element was taken. 
\begin{algorithm}[]
 \KwData{Runs to merge $R_1...R_k$, min-heap $H$, result run S}
 \KwResult{Data from runs to merge are in key-order in S, with the latest value corresponding to each key.}
 \For{each run $r$ 1...k}{
 push($H$, ($R[r][0], r$))\;
 }
 j := -1\;
 $lastKey$ := None\;
 $lastK$ := None\;
 $Heads$ := Array[k]\;
 $Heads[i]$ := 0 for all i \;
 \While{size($H$) $> 0$}{
  $(e, k)$ := pop($H$)\; 
  \eIf {e.key $==$ lastKey} {
  \eIf {lastK $<$ k}{
  S[j] = e\;
  } {}
  } {
  $j = j + 1$\;
  S[j] = e\;
  }
  lastKey = e.key\;
  lastK = k\;
  \eIf{Heads[k] $<$ size(R[k])}{
  $Heads[k]$ = $Heads[k] + 1$\;
  push($H$, $R[k][Heads[k]]$)\;
  }{}
  Construct-Index(S)\;

  }
 \caption{HeapMerge}
\end{algorithm}

The heap disgorges key-value pairs in key order; this entails that the result will be sorted, and in the case of multiple values for the same keys, only the highest-ranked run's value is written into the result. Though we do not include it in the formal algorithm specification, we mention that our implementation involves knowledge as to whether the merge in progress is writing to a new level below all the others. In that case, keys flagged for delete are not written to the buffer at all. The time and space bounds are trivially derived by noticing that $n$ elements are popped off the heap, and the heap is of size $mD$. In our implementation, we also use multithreaded merging to decrease our latency. When an insertion triggers a merge, a dedicated merge thread takes ownership of the runs to merge and executes the merge in parallel, allowing the main thread to rebuild the buffer and continue to answer queries. If a lookup request comes while the merge thread is executing the merge, the main thread searches the memory buffer for the requested key, and if unsuccessful, waits for the merge to complete before querying the disk levels.
\subsection{Insertion}
The algorithm for insertion into the sLSM is given in Algorithm 2. 
\begin{algorithm}[]
 \KwData{key k, value v to insert, runs runs[] of sLSM, Bloom filters B[] of sLSM, active run index $r_a$, size of runs $R_n$, number of runs $R$}
 \KwResult{key-value pair is inserted into the sLSM}
 \eIf{size(runs[$r_a$]) == $R_n$}{$r_a = r_a + 1$
 }
 {}
 \eIf{$r_a == R$}{Do-Merge(1)\;
 $r_a = r_a - mR$\;
 }{}
 insert(runs[$r_a$], k, v)\;
 insert(B[$r_a$], k, v)\;
  \caption{Put}
\end{algorithm}
The Do-Merge algorithm takes one parameter, which represents the disk level to merge runs to. In our implementation, it is a recursive function that merges successive levels until there is a free run to merge to, or it creates a new level at the bottom of the sLSM. Do-Merge calls HeapMerge when it finds an empty run to merge to, or after it creates a new, empty level. It is at this point that HeapMerge receives information as to whether the merged level is the last, in which case deletes are `'committed", i.e., not written to disk. Expected insertion time is calculated as follows: with probability $\frac{R_nRm -1}{R_nRm}$ insertion is into a skiplist with no merge necessary, i.e., $O(\log R_n)$. With probability $\frac{1}{R_nRm}$, a merge is necessary. Merging $R_nRm$ elements to disk level 1 takes $O(R_nRm (1 - \log \epsilon))$ time, given the necessity to copy each element for each element in the merged runs, as well as to write a Bloom filter (which requires $O(-n\log \epsilon)$ time for $n$ elements). With probability $\frac{1}{R_nRmD}$, the current insertion is the one that necessitated the first level to merge down to level two, which is an $O(R_nRm^2D \log Dm)$ operation. By induction, the probability that the current insert causes a merge down to level $k$ is $\frac{1}{R_nRm^{k+1}D^k}$, meaning that the total insertion time simplifies as 
\begin{eqnarray*}
&\frac{R_nRm -1}{R_nRm} O(\log R_n) + \sum_{k = 0}^L  \frac{R_nRm^{k+1}D^k(1-\log \epsilon) \log Dm}{R_nRm^{k+1}D^k}\\
&= O(\log R_n + (1-\log \epsilon) L \log Dm)
\end{eqnarray*}
with $L = O(\log n)$, where $n$ is the number of elements in the skiplist; this is due to the fact that levels are added at exponentially increasing thresholds of $n$.
\subsection{Lookup}
Lookups all follow the same pattern: Starting with the memory buffer and moving downwards to disk levels, query runs in newest-to-oldest order. For each run query, check if the key is between the min and max key for that run. If so, query the Bloom filter. If positive, then search the run. For in-memory runs, this entails searching the skiplist. For disk runs, this involves a binary search of fence pointers in order to find two file locations that bound the key, if it exists. Then a binary search is performed between those two locations and the key's value, if any, is returned. The worst case for a lookup is that the key doesn't exist, meaning that the algorithm has to search each level and each run. Noting that the time to query a Bloom filter with $k = - \frac{\log \epsilon}{\log 2}$ hash functions is $O(-\log \epsilon)$, the expected runtime of a lookup is
\begin{equation*}
O\bigg[(- \epsilon \log \epsilon)\bigg((R \log R_n)\ +
\sum_l^L \sum_k^D \log \left(\frac{R_nRm^kD^k}{\mu}\right) + \log \mu \bigg)\bigg]
\end{equation*}
where $\mu$ is the number of elements per fence pointer. This expression simplifies to
\begin{equation*}
O((-\epsilon \log \epsilon) (R\log R_n + DL\log R_n + DL\log R + D^2L \log Dm))
\end{equation*}
In practice, $R_n >> R$, so average lookup time is approximately $$O\left(\left(-\epsilon \log \epsilon\right) \left(\left(R+DL\right) \log R_n + D^2L \log \left(Dm\right)\right)\right)$$.
\subsection{Delete}
Deletes are implemented quite simply: to delete a key, simply insert that key paired with a value that signifies a deleted key. When a lookup comes across this value for this key, it immediately returns with a failure to find the key. Last, when this deleted key-value pair is merged down to start a new deepest level, it is omitted from the result set, as it is no longer needed to supersede any previous keys.
\subsection{Range}
Range queries involve looking up, for each run, all the elements in the range. For skiplists, this is as simple as locating the node corresponding to the smallest key greater than or equal to the first key in the range. Then, simply follow the skiplist's pointers until the current node is greater than or equal to the second key in the range, or else the end of the list has been reached. For disk-based runs, we first filter by key, then do only the 1 or 2 lookups we need to find the indexes in the run that frame the range. From there, we construct a hash table as follows: Starting with the newest run and working backwards towards the oldest, find all elements in the range, and for each, 1) insert the key and value into the hash table and 2) if the element is not a delete or already in the table, write it to the result set. The hash table guarantees that only the newest non-deleted values will remain in the result set. For our hash table, we again use the Murmur3 hash function. We use linear probing rather than chaining to optimize for small key-value pairs (as the test workload will be integers), and keep true key-value objects in the table, rather than pointers, in order to remain cache-optimal. When the hash table is more than half full, we double its size and rehash each element, leading to amortized $O(1)$ insertion and true constant-time probing. Because collecting elements in the range in both skiplists and disk runs is a linear-time operation, and because hashing is an amortized constant-time operation, a range query over $n$ keys is expected to take $O(n)$ time.
\subsection{Parameter table}
The full range of tuning parameters is given in Table 1.

\begin{table}[]
\centering
\caption{Parameters and Values}
\begin{tabular}{|l|l|l|}
\hline
\textbf{Parm} & \textbf{Meaning} & \textbf{Range} \\ \hline
$R$ & Number of runs & $\mathbb{Z} > 0$ \\ \hline
$R_n$ & Elements per run & $\mathbb{Z} > 0$ \\ \hline
$\epsilon$ & Bloom filter FP rate & (0, 1) \\ \hline
$D$ & Number of disk runs per level & $\mathbb{Z} > 0$ \\ \hline
$m$ & Fraction of runs merged & (0, 1] \\ \hline
$\mu$ & Fence pointer page size & $\mathbb{Z} > 0$ \\ \hline
\end{tabular}%
\end{table}
\subsection{Object Orientation}
We designed sLSM to be fully general; for the sake of academic experimentation, it makes sense to be able to substitute key and value data types at will, as well as to swap out run types. To this end, we chose C++ as our language, primarily for its speed and templating flexibility. With careful considerations such as a Run interface that enforces properties of a memory buffer run, we will in the future be able to simplify complex testing that involves trying different combinations of runs (perhaps hash tables or radix trees as well as skiplists).

\section{Experimentation}
We tested the sLSM on a DigitalOcean Droplet Server running 64-bit Ubuntu 4.4.0 with 32 Intel Xeon E5-2650L v3 @ 1.80GHz CPUs, a 500 GB SSD, 224GB main memory, and 30MB L3 cache.\\
It was clear that the choice of parameters for the sLSM would play an important role in performance optimization. Our first task was to find the combination of parameters that resulted in excellent baseline performance. Then, for various experiments, we deviated from that set of parameters in order to determine the effect of each parameter on overall performance in the face of changing workload types. To find this baseline combination of parameters, we tested on the Cartesian product of the parameters in Table 1 and data size up to 100MB, essentially performing a fine-mesh multidimensional grid search. Of the 1,556 parameter sets that we sampled, we chose the top 10\% to move on to the next round of larger data set testing. We then tested each one of the remaining parameter sets on 500MB, 1GB, and 10GB data sets, selecting the parameter set with the highest average (weighted by number of inserts/lookups) insert plus lookups per second. This baseline parameter set is the basis for the rest of our experimentation: $\mu = 512, \epsilon = 0.001, R = 50, R_n = 800, D = 20, m = 1.0$. Unless otherwise specified, all experiments use these parameters, as well as a 100 million key dataset with 32-bit integer keys generated uniformly at random. 
\subsection{Number of Runs in Memory}
In determining the optimal $R$, we found that the smaller $R$ is, the smaller the memory buffer is, and the more frequent merges will be. Thus, lower $R$ leads to lower insertion throughput. However, with few runs to search, lookups are very quick with small $R$. Analogously, higher $R$ is linked to faster insertion but slower lookup, since more runs need to be searched. With Bloom filters, it is possible to set $R$ high enough to achieve extremely fast insertion while enjoying significant speedup on lookups due to the filters. More formally, $R$ does not enter the amortized insertion time function, and there are significant constant factors hidden in that equation that correspond to the speed and frequency of merges. However, lookups depend linearly upon $R$, as proven above.
\begin{figure}
\centering
\includegraphics[scale=.8]{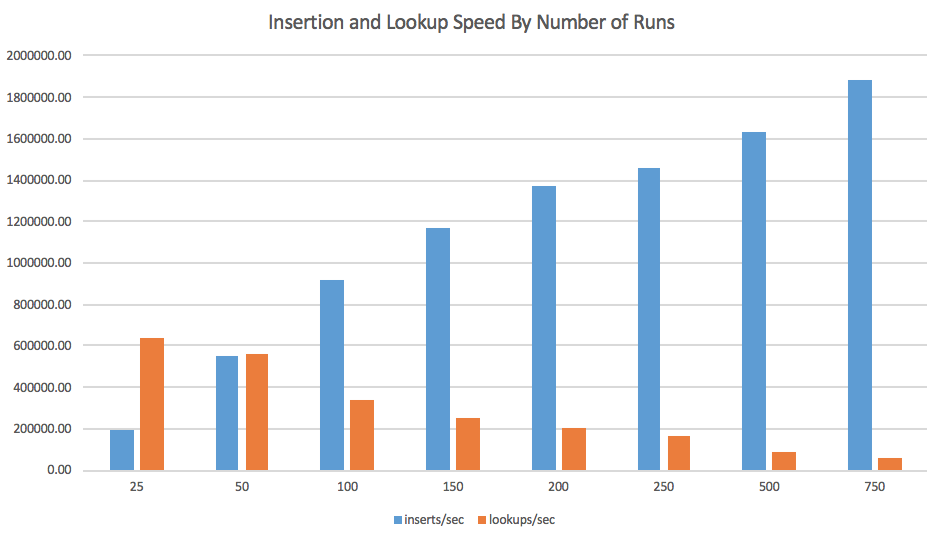}
\caption{}
\label{numruns}
\end{figure}
The graph in Fig. \ref{numruns} details the tradeoff between insertion time and lookup time for a number of values of $R$. As such, setting the number of runs intelligently also allows us to tune the performance of the sLSM to the workload at hand- more runs for more writes, and fewer for more lookups.
\subsection{Buffer Size}
\begin{figure}
\centering
\includegraphics[scale=.8]{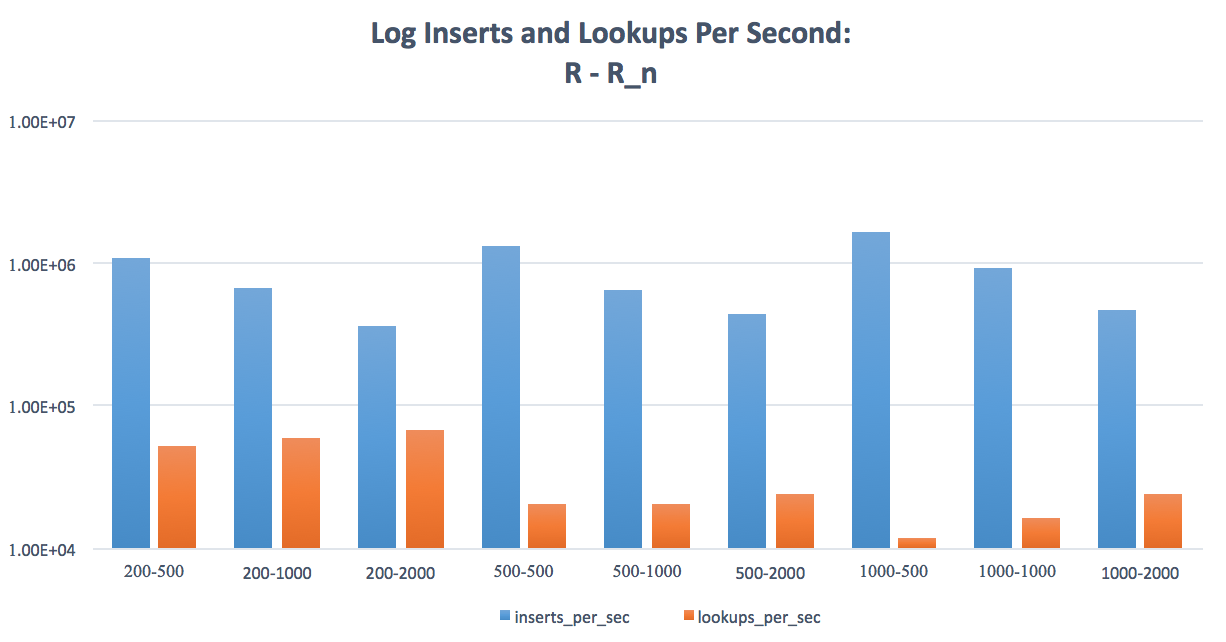}
\caption{}
\label{bufsize}
\end{figure}

The choice of buffer size is tightly linked with the experimental optimization of the number of runs. However, instead of a simple size-in-bytes parameter for the memory buffer, we expose a two-dimensional knob. We allow the user to choose both the number of runs and the size of each run. As shown in Fig. \ref{bufsize}, the number of runs is crucial for performance. However, another important factor for speed is the size of each skiplist. As shown, $R_n$ is a main determinant of the insertion and lookup speed within each run. For this experiment, we take a Cartesian product over $R \times R_n$. As expected, insert and lookup throughputs are traded off as $R$ increases. However, of interest here is the effect of changing $R_n$. For each value of $R$, increasing $R_n$ increases insertion rate while decreasing lookup throughput. This is because a larger $R_n$ allows for fewer merges over the lifetime of the workload due to the larger memory buffer. However, this increase in the size of each run increases the runtime of each lookup, since skiplist queries are logarithmic in their size. For this reason, we chose a value of $R_n$ between 500 and 1000, finding that 800 worked nicely for a variety of workloads.
\subsection{Disk and Merge Parameters}
\begin{figure}
\centering
\includegraphics[scale=.8]{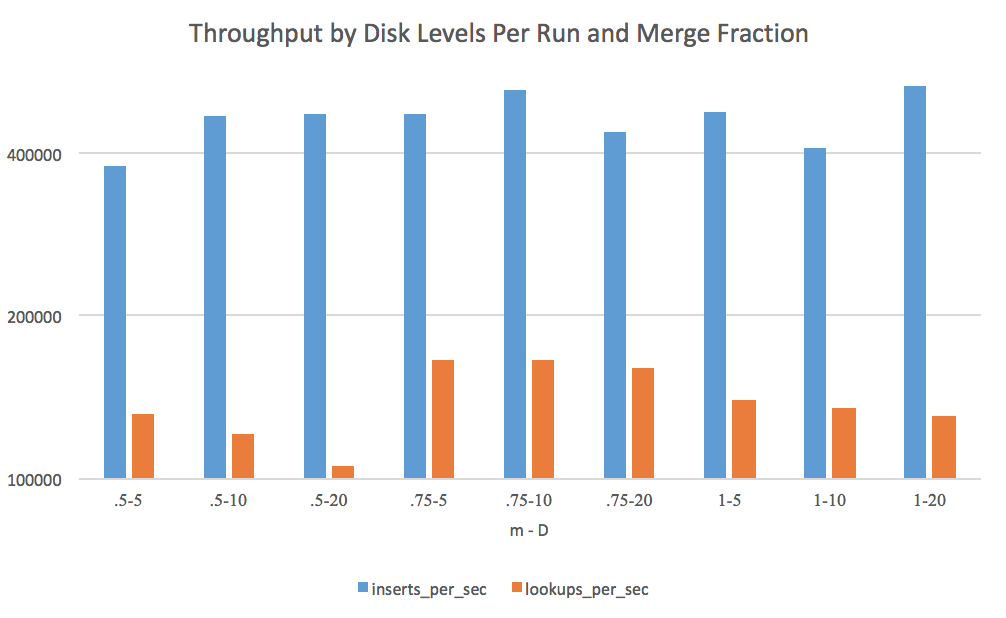}
\caption{}
\label{md}
\end{figure}

We determined that if $m$ is set under $0.5$, merges would happen too frequently for sizeable datasets, causing the OS to run out of file descriptors. To determine optimal values of $D$ and $m$, we took their Cartesian product for along the range of values that were left after the original parameter tuning. For our standard workload, there was not a significant trend other than that throughput for lookups tended to decrease as $D$ increased for small $m$. This is likely due to the fact that the OS had too have many files open, leading to page churn as very small disk runs are searched in sequence. We present our chart with these results in Fig. \ref{md}, noting that even if the experimental results are sorted by $Dm$, the size ratio between levels, there is not a significant trend. Further experimentation will involve larger datasets given this Cartesian product, as well as the introduction of various types of workloads.\\
\subsection{Bloom Filters}
Here we describe the performance enhancements afforded by Bloom Filters. Testing on a workload of a million inserts and lookups, we demonstrate the following cases: no Bloom filter; Bloom filter with $\epsilon \in \{0.1;0.01;0.001;0.0001;0.00001;0.000001\}$\\
\begin{figure}
\centering
\includegraphics[scale=.8]{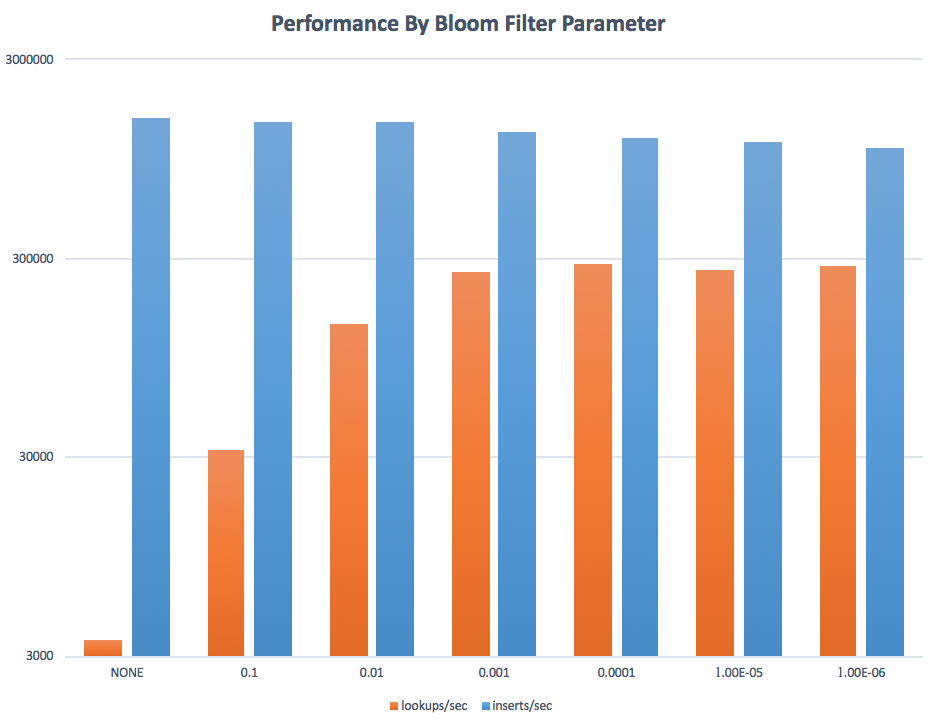}\\
\caption{}
\label{bf}
\end{figure}
As shown in Fig. \ref{bf}, the filter provides an impressive speedup, from 3,634 lookups/sec to over 340,000/sec, with no significant difference in insertion time. The intuition behind the speedup is simple: for each run, we avoid a lookup inside it if we fail a Bloom filter test. Since such a test is far cheaper than a skiplist lookup, we save ourselves the time of doing the deep search by ruling out the possibility that a key exists in a particular run.\\Under profiling, we found that 98.9\% of the CPU (clock) time is spent in the skiplist lookup function without Bloom filters. This drops to a mere 13.1\% when the filters are introduced. In that case, 53.2\% of the CPU time is spent calculating the hash functions necessary for the filter. Perhaps there is an opportunity for further optimization here; finding quicker hash functions or making the filter structure more cache-conscious could ostensibly provide significant speedup, since the filters are such "hot" structures. A more in-depth optimization might take the form of recent work by Dayan et al. in which the false positive rates of the Bloom filters are dynamically optimized across different levels of the structure\cite{521291}.
\subsection{Range Queries}
\begin{figure}
\centering
\includegraphics[scale=.7]{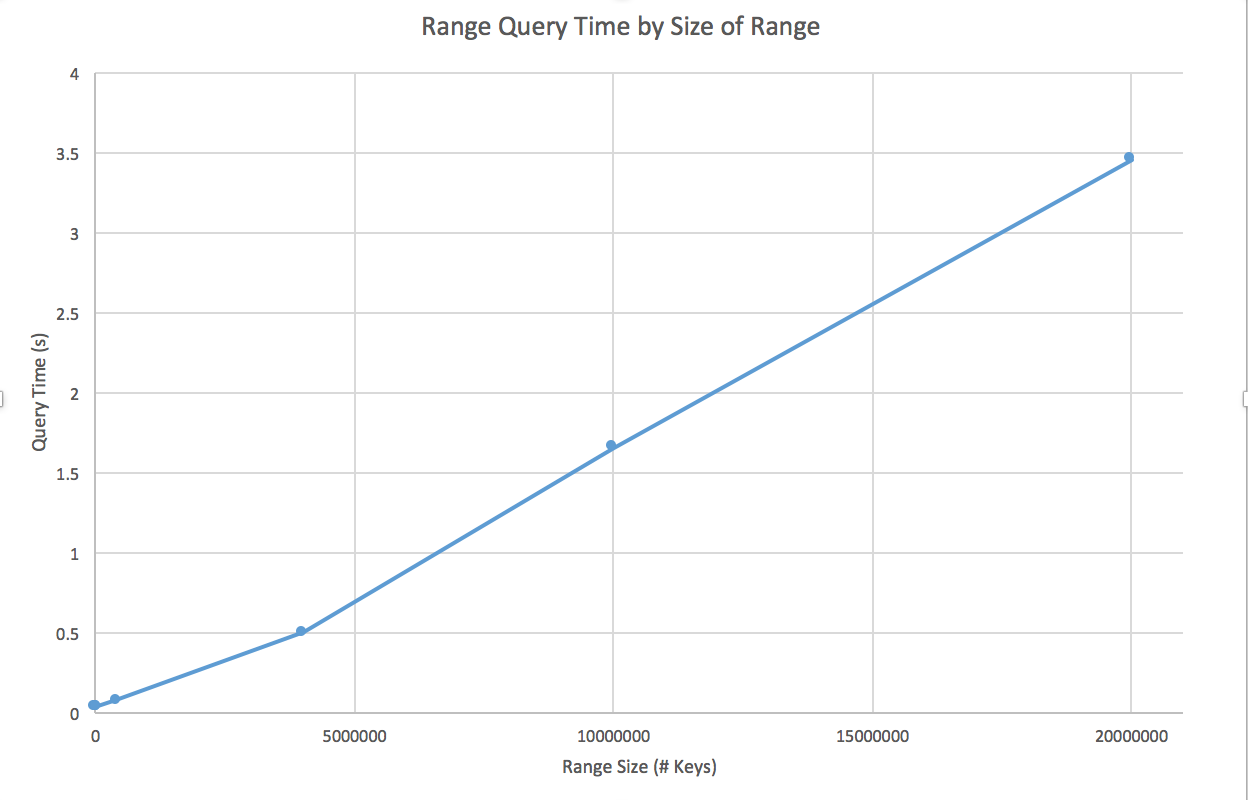}
\caption{}
\label{range}
\end{figure}

We present a short demonstration of the sLSM's range query performance. As derived in Section 2, range queries are linear time operations in the size of the range. For several different range sizes, we plot the time to complete range queries over a uniform distribution of keys in Fig \ref{range}.\\
\subsection{Data Size}
We now show results for varying dataset sizes. As is evident, insertions into and lookups experience slowdowns as the data size gets larger.\\
\begin{figure}
\includegraphics[scale=.8]{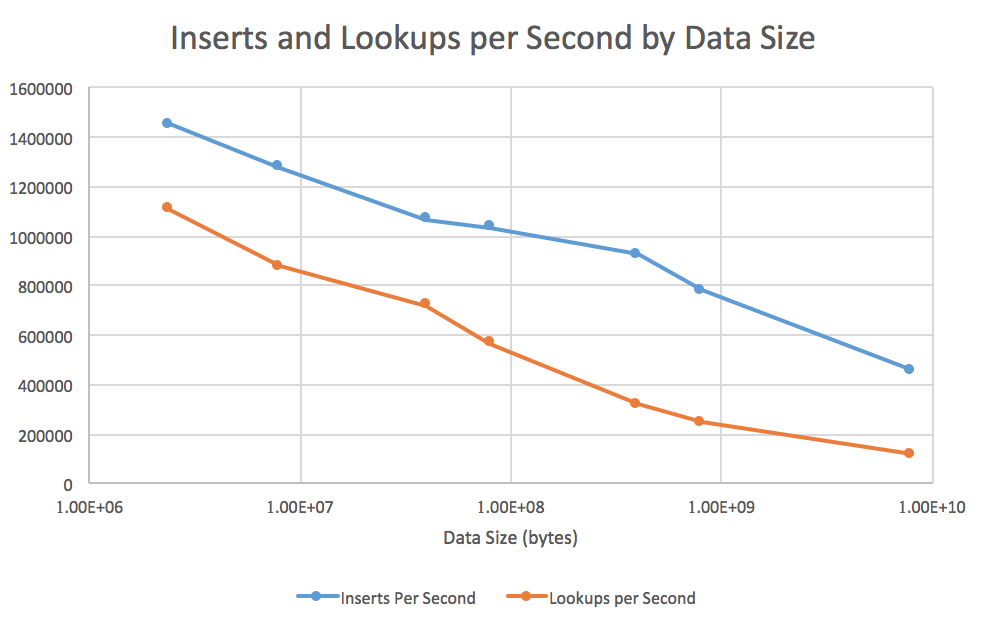}\\
\caption{}
\label{size}
\end{figure}
\subsection{Data Size}
The reason for this effect with respect to inserts is that as the system gets larger, it completes more merges, which require expensive disk accesses. Lookups too require querying more Bloom filters and disk runs as the data grow large, meaning that each lookup requires more time. Nevertheless, our system exhibits no more than a logarithmic slowdown, which is as good as it can be given the theoretical results shown above. See Fig. \ref{size} for details.
\subsection{Update-Lookup Ratio}
\begin{figure}
\centering
\includegraphics[scale=.8]{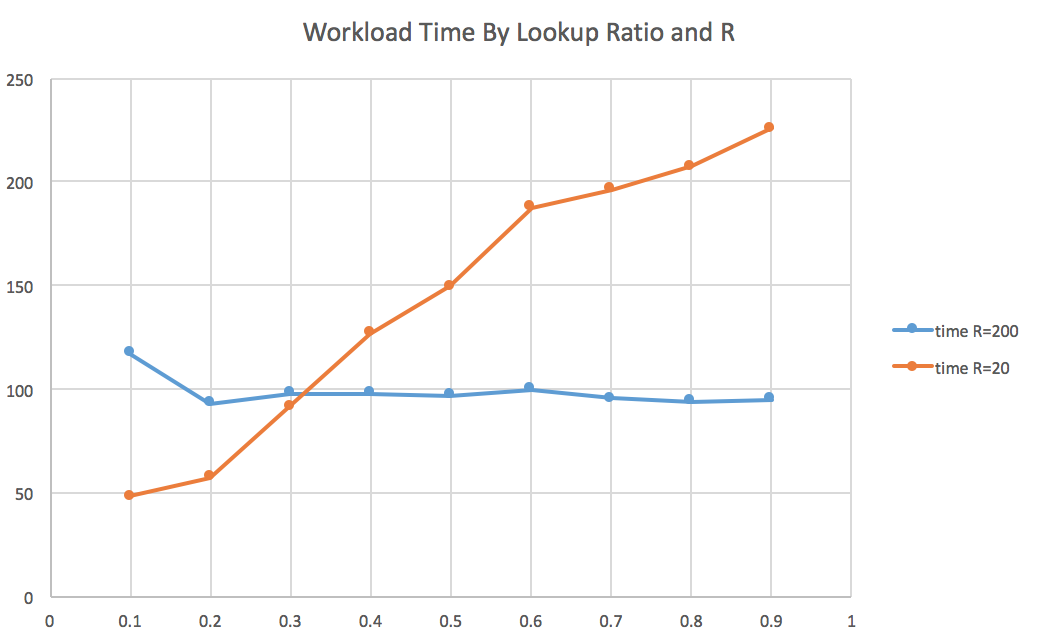}
\caption{}
\label{ulr}
\end{figure}

For many data systems, performance is dependent upon the ratio of updates to lookups in the workload. In this experiment, we manipulated this ratio between 10\% lookups and 90\% lookups for a 100 million query workload. To demonstrate the ability of sLSM to adapt to various workloads, in Fig. \ref{ulr} we show plots of completion times of the query set for two sLSMs: one parameterized by $R = 20$ and one by $R = 200$. As shown above, higher $R$ leads to increased insert throughput at the cost of lookup speed. As such, the graph displays that the tree with balanced parameters ($R = 20$) is quite forgiving with respect to the lookup ratio. In contrast, the specialized tree ($R = 200$) completes the low-lookup, high-insert workloads an order of magnitude quicker than its balanced counterpart, at the cost of steeply decreasing performance as lookups become more prevalent in the workloads.\\
\subsection{Data Skew}
\subsubsection{Insertion Skew}
\begin{figure}
\centering
\includegraphics[scale=.8]{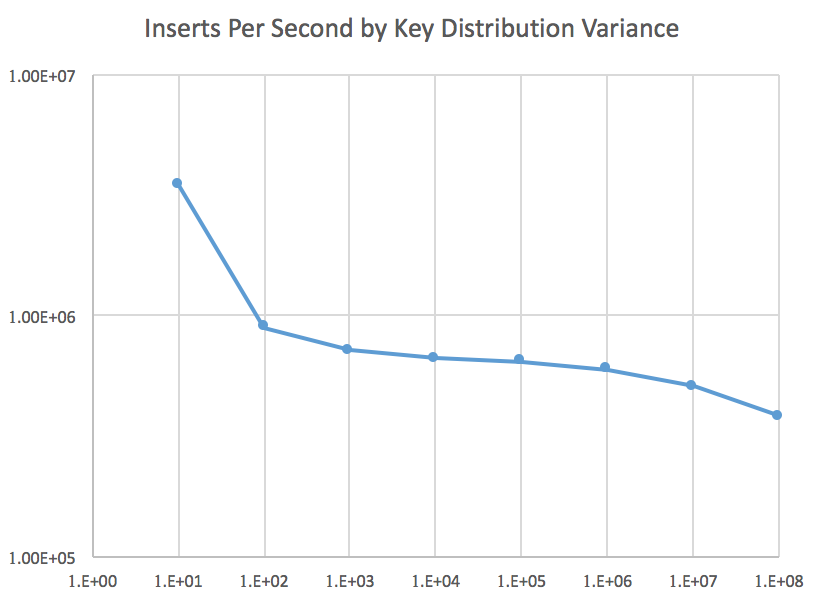}
\caption{}
\label{insertskew}
\end{figure}

The skewness of inserted data is a vitally important factor of performance. Our skiplists do not blindly insert elements; rather, if a key is already in the active skiplist, its value is simply updated. This means that for data with low variance of keys, insertion can be incredibly fast. For this experiment, we generate integral keys via a normal distribution (rounding to the nearest integer) around zero and manipulate the variance. As the variance increases, there is greater variety of keys, and the insert performance drops precipitously, as shown in Fig. \ref{insertskew}.\\
\subsubsection{Lookup Skew: Single Threaded}
\begin{figure}
\centering
\includegraphics[scale=.8]{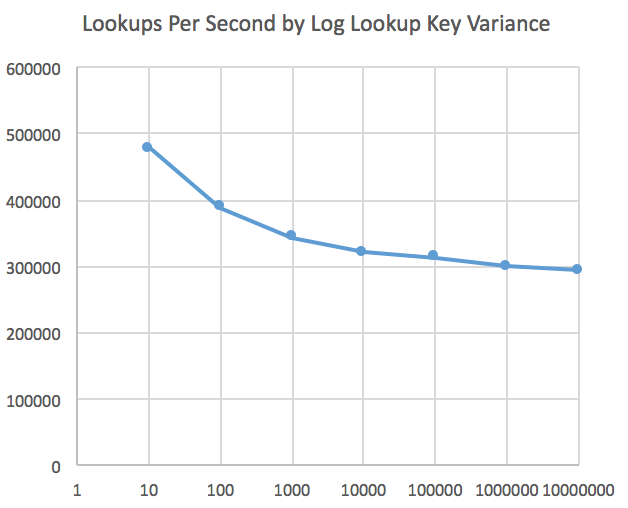}\\
\caption{}
\label{lookupskewsingle}
\end{figure}

Lookup skew is similar: in the same style of experiment, we find that for a set of uniformly distributed keys in the sLSM, querying for a tightly clustered set of lookup keys results in higher performance. The performance degrades as the lookup keys get more dispersed. A small set of lookup keys requires fewer random seeks and disk page loads, resulting in better performance, as shown in Fig. \ref{lookupskewsingle}. Further work will involve testing on hard disk drives and on systems that allow for fine-grained tracking of disk IOPS.\\
\subsection{Concurrency}
\subsubsection{Lookup Skew: Multithreaded}
\begin{figure}
\centering
\includegraphics[scale=.8]{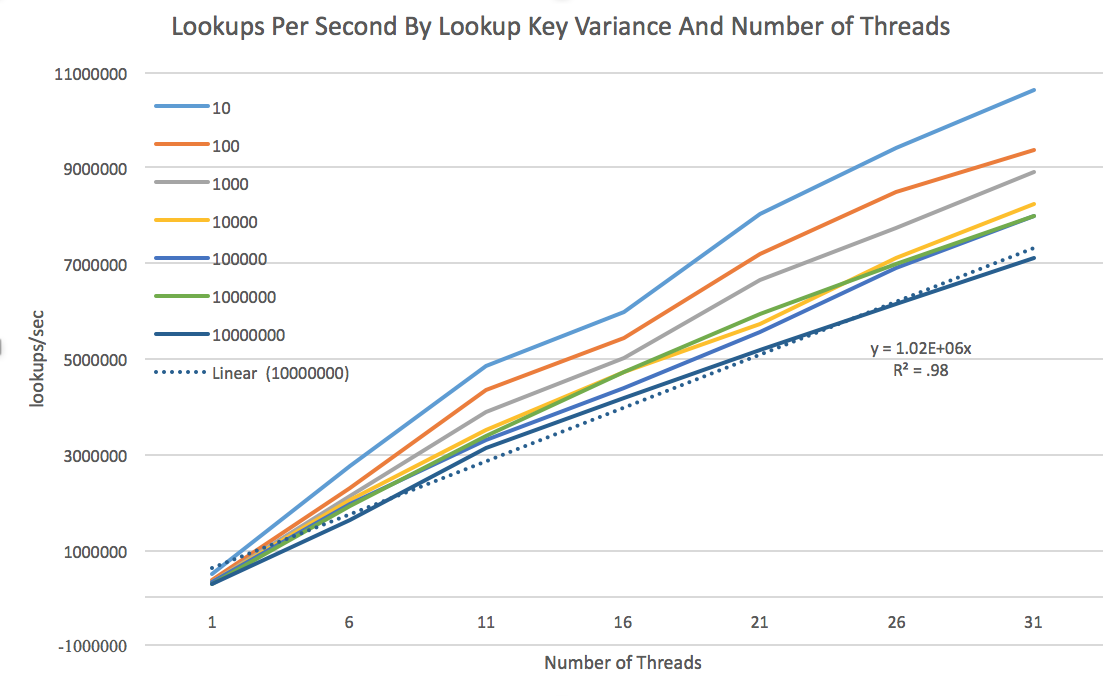}\\
\caption{}
\label{lookupskewmulti}
\end{figure}

Lookup skew also provides an opportunity to utilize concurrency in our experimentation. In this experiment, lookup skew was varied along with the number of threads performing concurrent lookups, demonstrating that with highly clustered lookups, the sLSM's scaling factor is higher with each thread than with evenly-distributed lookups. This is due to the fact that the disk is able to optimize its seeking to service the lookup requests better when there is a small locality of keys. Perhaps multiple threads could even be serviced by the same disk pages, cutting down on disk operations even further. The closeness of the keys allows the access pattern to act like sequential requests rather than random requests. The graphic in Fig. \ref{lookupskewmulti} also highlights the way that the sLSM scales with the number of threads in the general (key-dispersed) case. With a correlation coefficient of $R^2 = 0.98$ and an increase in lookup throughput of $1.02$M per thread, sLSM proves its worth as a parallel data structure as well.\\
\subsubsection{Merge Threading}
One of the concurrency optimizations we implemented was the dedication of a hardware thread for merging disk levels to allow for lower latency. For this test, we measured the largest time between insertions over 100M keys. We used our server's SSD as well as a Toshiba Canvio HDTB205XK3AA 500GB external hard drive connected via USB3.0. With this setup, we were able to measure the reduction in tail latency afforded by merge threading for both spinning and solid-state disks.\\
\begin{figure}
\centering
\includegraphics[scale=.8]{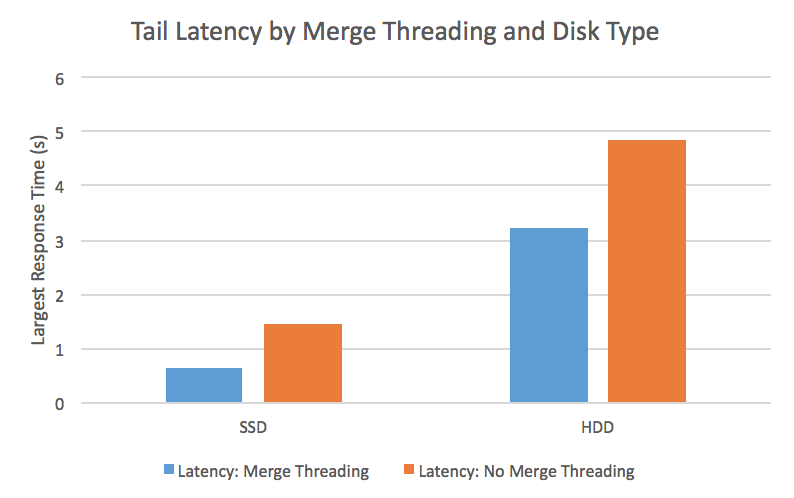}\\
\caption{}
\label{mergethread}
\end{figure}
As shown in Fig. \ref{mergethread}, there is a significant reduction in maximal response time for both SSD and HDD. In further testing, we also showed that merge threading allowed the system to experience no less than 99.7\% CPU utilization throughout the workload, while utilization dropped to 53\% at times without merge threading due to the processor waiting for the disk operations to finish.
\section{Conclusion}
We presented the Skiplist-Based Log Structured Merge Tree, a data system that makes heavy use of probabilistic data structures to provide highly performant transactional queries. Our novel approach includes the development of a sequence of cache-aware skiplists, indexing via Bloom filters and fence pointers, a fast k-way merging algorithm, lookup and merging concurrency, and a thorough experimental evaluation that details the tradeoffs between update and lookup throughput for a wide variety of workloads.\\We also showed theoretical guarantees as to the performance of the system at scale and corroborated them  empirically, demonstrating that the system's performance is tightly bounded by the theoretical guarantee. Further, we demonstrated that the sLSM is adaptable to various workloads, readily able to perform well for query sets of different types of skew. In single threaded work and with the right tuning, the sLSM can exceed millions of queries per second for both updates and reads, far outpacing baseline results for such systems. Concurrent lookups scale nearly perfectly as well, allowing the sLSM to achieve throughput of between 7 and 11 million lookups per second on various datasets.\\While there is still testing and implementation work to be done, the sLSM has proven its worth as a subject of research and as a high-performance big data system.